\documentclass[twocolumn,prl,aps,unsortedaddress,showpacs]{revtex4-1}
\usepackage{url,psfrag,graphicx}
\usepackage{dcolumn}
\usepackage{amsmath,amssymb}
\usepackage{bm}
\usepackage{pstricks,multido}
\usepackage{hyperref}
\usepackage{epsfig}
\def\<{\left<}
\def\>{\right>}
\def\ket|#1>{\left|#1\right>}
\def\bra<#1|{\left<#1\right|}
\def\elem<#1|#2|#3>{\left<#1\right|#2\left|#3\right>}
\def\braket<#1|#2>{\langle#1|#2\rangle}
\def\({\left(}
\def\){\right)}

\psset{unit=1mm}

\begin{document}

\title[Short Title]{Energy space entanglement spectrum of pairing models
with $s$-wave and $p$-wave symmetry}

\author{Javier Rodr\'{\i}guez-Laguna}
\affiliation{Mathematics Dept., Universidad Carlos III de Madrid, Spain}
\affiliation{Institute of Photonic Sciences (ICFO), Barcelona, Spain}
\author{Miguel Ib\'a\~nez Berganza}
\affiliation{Istituto per i Processi Chimico-Fisici (CNR) and
Dipartimento di Fisica, Universit\`a ``La Sapienza'', Roma, Italy}
\author{Germ\'an Sierra}
\affiliation{Instituto de F\'{i}sica Te\'orica (IFT), UAM-CSIC, Madrid, Spain}

\date{July 31, 2013}

\begin{abstract}
Entanglement between blocks of energy-levels is analysed for systems
exhibiting $s$-wave and $p$-wave superconductivity. We study the
entanglement entropy and spectrum of a block of $\ell$ levels around
the Fermi point, and also between particles and holes, in the ground
state of Richardson-type Hamiltonians. The maximal entropy grows with
the number of levels $L$ approximately as $1/2\log(L)$, as suggested
by the permutational symmetry of the state at large coupling. The
number of levels in the block around the Fermi surface with maximal
entanglement is proposed as a measure of the number of {\em active
Cooper pairs}, which correlates with standard estimates of this
magnitude. The entanglement spectrum is always composed of a principal
parabolic band plus {\em higher bands} whose disappearance signals a
exact BCS state, e.g. in the Moore-Read line, while the Read-Green
quantum phase transition is characterized by a maximum in their
weight.
\end{abstract}


\maketitle



Entanglement has proved to be a extremely useful concept in order to
characterize quantum phase transitions (QPT) \cite{Amico.RMP.08}. The
usual objects of study are the various {\em entanglement entropies}
between a real space block and the rest of the system. In 2008, Li and
Haldane showed that the {\em entanglement spectrum} (ES) was much more
informative \cite{Li_Haldane.PRL.08}. In particular, these authors
found that the {\em low energy} ES of the $\nu= 5/2$ fractional
quantum Hall (FQH) states is described by the chiral Conformal Field
Theory associated to the edge excitations. This is an example of the
so called bulk-edge correspondence which has been explored intensively
in the last few years not only in the FQH context but also in spin
systems, topological states, random systems, etc.  \cite{ent_es}.

The previous works concern mostly the characterization of entanglement
in real space, or for the Hall systems in the orbital basis. However,
some physical systems are formulated in momentum space or in energy
space. A notable example is provided by superconductors where the
Hilbert space is based on {\em single-particle energy levels}. In that
spirit, Martin-Delgado computed the {\em concurrence} between modes in
the grand canonical BCS state \cite{MAMD.02}, while Dunning and
collaborators computed the concurrence in the canonical ensemble using
the exact solution of the BCS model \cite{Dunning.PRL.05}.
Entanglement between regions in {\em Fourier space} has been studied
in \cite{Balasubramanian.PRD.12}. Moreover, successful application of
the density matrix renormalization group (DMRG) techniques in momentum
space \cite{Xiang.PRB.96} suggests that entanglement in Fourier space
might be significantly lower than in real space for some problems.

In this work we study the entanglement spectrum and entropy between
disjoint blocks of single-body energy levels in one-dimensional models
of fermion pairing presenting both $s$-wave and $p$-wave symmetry. The
analysis is carried out in the canonical ensemble, assuming that
fermion pairs behave like hard-core bosons \cite{Dukelsky.RMP.04}. As
our main numerical technique, we employ the DMRG \cite{White.PRL.92},
which has already proved useful in those systems
\cite{Dukelsky_Sierra.PRL.99,Dukelsky_Sierra.PRB.00}.

 
Let us consider an ultrasmall superconducting grain with $L$ energy
levels and $N_e=2N_p$ electrons subject to an $s$-wave pairing
interaction. Because of its irregular shape, momentum is not
preserved, but single-particle states come in time-reversed pairs,
denoted as $\ket|j+>$ and $\ket|j->$, with the same energy
$\epsilon_j$. Let $b^\dagger_j$ be a hard-core boson operator which
creates that pair. As usual in BCS systems, let us assume a
homogeneous pairing coupling among all energy levels, with coupling
constant $G$. By restricting ourselves to states in which each energy
level is either empty or doubly occupied, we obtain the following
Hamiltonian, known as {\em Richardson model}
\cite{Dukelsky.RMP.04,Richardson.PL.63,Richardson.JMP.65,Gaudin.95}:
$H_R=\sum_{j=1}^L 2\epsilon_j b^\dagger_j b_j - G \sum_{j\neq j'}^L
b^\dagger_jb_{j'}$. We will work with equally spaced levels,
$\epsilon_j=jd$, with $j\in \{1,\cdots,L\}$. This choice can be
heuristically justified based on the level repulsion in random systems
\cite{Sierra.PRB.00}. Making $d=1/L$ and defining an effective
coupling constant $g=G/d=GL$, we get

\begin{equation}
H_s/d= \sum_j 2j\ b^\dagger_j b_j - g \sum_{j\neq j'}
b^\dagger_jb_{j'}.
\label{ham.richardson}
\end{equation}

For $g=0$, the ground state of the Hamiltonian is the Fermi state with
$N_p$ pairs $\ket|\Psi_0>=\prod_{k=1}^{N_p} b^\dagger_k \ket|0>$. As
$g$ increases, more pairs jump to excited states, and for
$g\to\infty$, all the energy levels are equally occupied and the state
becomes symmetric under permutations.

We have employed the DMRG to obtain numerically the ground state of
the Hamiltonian (\ref{ham.richardson}) for different numbers of levels
$L$, pairs $N_p$ (filling factor $x=N_p/L$) and coupling constant
$g$. The DMRG was shown to provide excellent results by Dukelsky et
al. \cite{Dukelsky_Sierra.PRL.99}, using an adapted version of the
infinite algorithm \cite{White.PRL.92}. We have extended their methods
by employing the finite DMRG algorithm with a number of retained
states adapted to keep the sum of the neglected weights always below
$10^{-8}$.

\begin{figure}
\epsfig{file=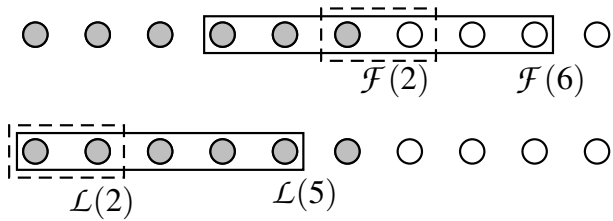,width=9cm}
\caption{\label{fig.blocks}Illustration of the two types of blocks
  whose entanglement we study.}
\end{figure}

It is well known that the low energy physics of the BCS systems takes
place in the vicinity of the Fermi level $\ell_F$. Thus, it makes
sense to study the entanglement of what can be called {\em Fermi
  blocks}: ${\cal F}(\ell)$ contains the $\ell$ levels which are
closest to $\ell_F$, i.e.: ${\cal
  F}(\ell)=\{\ell_F-\ell/2+1,\cdots,\ell_F+\ell/2\}$. Figure
(\ref{fig.blocks}, top) illustrates the definition. Notice that only
even values of $\ell$ are allowed in ${\cal F}(\ell)$ until it reaches
either the top or bottom of the energy scale. We shall also consider
{\em low-energy blocks}: ${\cal L}(\ell)=\{1,\cdots,\ell\}$ which
contain the $\ell$ lowest energy levels, as illustrated in figure
(\ref{fig.blocks}, bottom). In the particular case where $\ell=
\ell_F$ we recover the particle-hole partition used in reference
\cite{Dukelsky_Sierra.PRL.99,Dukelsky_Sierra.PRB.00} to implement the
DMRG for the Richardson model.

\begin{figure}
\epsfig{file=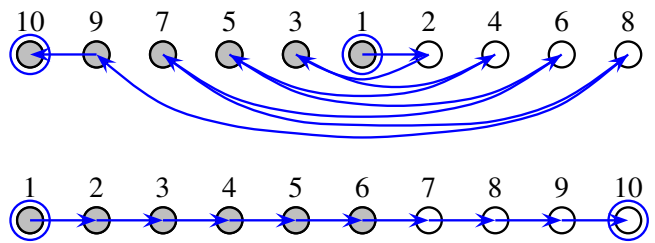,width=9cm}
\caption{\label{fig.paths}The two DMRG-paths used to compute the
  entanglement for each family of blocks.}
\end{figure}

It is rather straightforward to obtain the ES of the blocks ${\cal
  F}(\ell)$ and ${\cal L}(\ell)$, provided they correspond to the left
(or right) blocks of the DMRG. We have used two different DMRG-paths,
i.e.: two orderings of the sites, as illustrated in figure
(\ref{fig.paths}): the top panel shows the path used for Fermi blocks,
and the bottom one, the path for low-energy blocks.

\begin{figure}
\epsfig{file=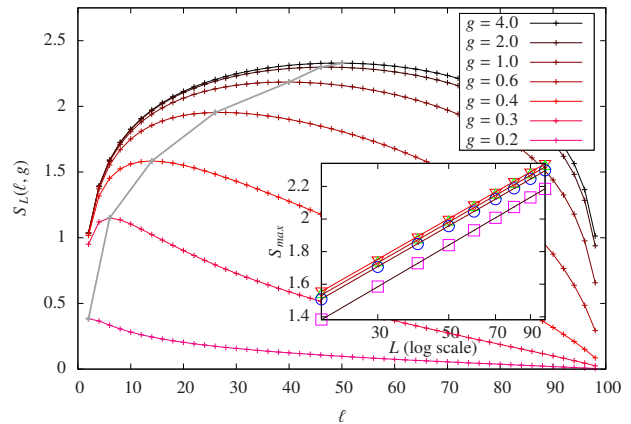,height=6cm}
\caption{\label{fig.swave.sl}Entanglement entropy $S_L(\ell,g)$, with
  maxima marked with polygonal line. Inset: $S_{max}(L,g)$ for $g=1$,
  $2$, $3$ and $\infty$ (increasing).}
\end{figure}

Let $S_L(\ell,g)$ denote the entanglement entropy (EE) associated to
the block ${\cal F}(\ell)$ in the GS of the Hamiltonian
(\ref{ham.richardson}) at half-filling, $x=N_p/L=1/2$. Figure
(\ref{fig.swave.sl}) shows its values for $L=100$ and several
couplings $g$. When $g\to 0$, the entropy vanishes, since the GS is a
Fermi sea with small fluctuations and is close to a product state. On
the other hand, when $g\to\infty$ the plot becomes symmetric around
$\ell=L/2$, since all the energy levels are equally occupied and the
state is symmetric under permutations.

As a function of $\ell$, $S_L(\ell,g)$ attains a maximum at a value
$\ell_{max}(L,g)$ which depends on $L$ and $g$, as shown in the gray
polygonal line in figure (\ref{fig.swave.sl}). As $g$ increases, this
maximum moves from $\ell_{max}=1$ to $\ell_{max}=L/2$ which is reached
at $g=\infty$. Let us denote the maximal entropy as
$S_{max}(L,g)\equiv S_L(\ell_{max},g)$. We show how it depends on $L$
and $g$ in figure (\ref{fig.swave.sl}, inset). As we will justify
later, for sufficiently large $g$ {\em or} $L$, we have:

\begin{equation}
S_{max}(L,g) \approx {1\over 2} \log(L) + \beta(g) + O(L^{-1})
\label{slogl}
\end{equation}

\noindent and the coefficient of $L^{-1}$ is $1/2$ for large
$g$. Entanglement entropies growing as a log of the block size have
been already observed in a variety of systems such as spin and fermion
chains at criticality, random 1D systems,
etc. \cite{Holzhey.NPB.94,Calabrese.JSTAT.04,Vidal.PRL.02,Refael.PRL.04,Laflorencie.PRB.05,Chen.13}.
The physical origin in our case is however different: it stems from
the permutation-symmetry that arises for $g\to\infty$, when the
entanglement of a block of $\ell$ levels only depends on its size. In
the strong coupling limit, the GS approaches quickly the state with
$S_z=0$ from the multiplet of a ferromagnetic state with $L$ spins
$1/2$, which in Quantum Information is known as the Dicke state at
half filling \cite{Stockton.03}. Several authors have shown that the
entanglement entropy of a block containing half the sites of those
states grows as $1/2 \log(L)$ \cite{Stockton.03,Popkov.05}. These
results have been generalized by Castro-Alvaredo and Doyon: the
entanglement entropy of states constructed by integrating degenerate
factorizable states grows as $d/2\log(L)$, where $d$ (in our case, 1)
stands for the {\em number of Goldstone bosons} associated to the
degeneracy \cite{CastroAlvaredo.12,Castro_Alvaredo.JSTAT.13}.

In the BCS model we can derive equation (\ref{slogl}) in a way that
also provides the entanglement spectrum. Following the analysis of
\cite{Dukelsky_Sierra.PRL.99}, we consider the decomposition of the
system, at half-filling, between the block $A={\cal L}(L/2)$ and its
complement, i.e.: particles vs. holes. This partition corresponds to
the Schmidt decomposition:
\begin{equation}
\ket|\Psi_L>=\sum_{n=0}^{L/2} \psi_{L,n}  \ket|n>_A\otimes \ket|L/2-n>_B
\label{richardson.gs.ginf}
\end{equation}
\noindent where $\ket|n>_A$ is a state with $n$ pairs below the Fermi
energy and $\ket|L/2-n>_B$ a state with $L/2-n$ pairs above it, with
$n=L/2$ corresponding to the Fermi state. Using a combinatorial
analysis one finds that $\psi_{L,n}=C_{L/2,n}/\sqrt{C_{L,L/2}}$, where
$C_{n,m}=n!/(m!(n-m)!)$ \cite{Dukelsky_Sierra.PRL.99}, and the
eigenvalues of the reduced density matrix of the block ${\cal L}(L/2)$
are given by $\lambda_{L,n} = |\psi_{L, n}|^2$.  In the limit $L, n
\gg 1$, the highest values of $\lambda_{L,n}$ are concentrated in the
vicinity of $n \sim L/4$. The entanglement energies, defined as
$\epsilon_{L,n} = - \log \lambda_{L,n}$, can be approximated by the
parabola
\begin{equation}
\epsilon_{L,n}  \simeq \frac{L}{2} 
\left( 1- \frac{ 4 n}{L} \right)^2 + \frac{1}{2} \log \frac{ \pi L}{8}, 
\quad n \sim \frac{L}{4}.
\label{epsi}
\end{equation}
so the most probable terms deviate from $n = L/4$ by an amount of
order $\sqrt{L}$. Extending the range of $n$ to the real line one
finds that (\ref{epsi}) is correctly normalized, i.e. $\int dn \, e^{-
  \epsilon_{L, n}} = 1$, and that the entanglement entropy, given by
$S= \int dn \, \epsilon_{L, n} e^{- \epsilon_{L, n}} = \frac{1}{2} (
\log ( \pi L/8) +1)$, reproduces the asymptotic behavior
(\ref{slogl}). Moreover, the next order Stirling approximation yields
a finite-size correction of $1/(2L)$, as expected. We have also
performed projected-BCS computations \cite{Dukelsky_Sierra.PRL.99},
obtaining again the form (\ref{slogl}) for large $g$ or $L$, thus
providing further support for its robustness.

\begin{figure}
\epsfig{file=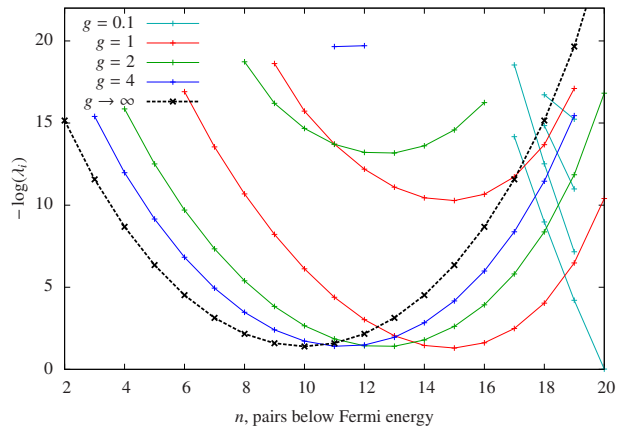,height=6cm}
\caption{\label{fig.swave.entsp}Entanglement energies, $\epsilon_i = -
  \log \lambda_i$ for the GS of (\ref{ham.richardson}). The dotted
  line corresponds to the limit $g\to\infty$ limit given approximately
  in eq.(\ref{epsi}).}
\end{figure}

For moderate values of $g$ one should expect that the particle-hole
Schmidt decomposition of the GS will involve more than one state
having the same number of particles (or holes), and not just one
state, as in eq.(\ref{richardson.gs.ginf}). This expectation is
confirmed by the numerical results. We find that the entanglement
spectrum is formed by parabolic bands that resemble the structure of
the principal band (\ref{epsi}), as shown in
Fig.(\ref{fig.swave.entsp}) for a system with $L=40$ energy levels and
$g=0.1$, $1$, $2$ and $4$. The entanglement energies $\epsilon_i = -
\log \lambda_i$ for the block ${\cal L}(L/2)$ are labeled by the
number of pairs below the Fermi energy. The dashed line depicts the
parabolic spectrum (\ref{epsi}) corresponding to the limit
$g\to\infty$. The band structure is clear. The vertex of each parabola
moves leftwards as $g$ increases: for $g\approx 0$ the maximal
probability corresponds to $n=\ell_F(=L/2)$, i.e.: the Fermi sea, all
pairs are below $\ell_F$. As $g$ increases, the minimum shifts
leftwards, approaching $n=\ell_F/2(=L/4)$ as $g\to\infty$: only half
the pairs, in average, are below $\ell_F$. As $g\to\infty$, the weight
of the {\em higher bands} in the EE is reduced. This implies that the
approximation given in equation (\ref{richardson.gs.ginf}), in which a
single weight is given to each occupation number, improves as the
coupling constant raises. An interesting question is what is the
entanglement Hamiltonian $H_E$ that gives rise to parabolic bands with
that structure.

Moreover, the study of entanglement enables us to estimate the size of
the {\em active region} in energy space. For a given value of the
coupling constant $g$, only a certain fraction of the energy levels
will be actively involved in the dynamics of the system. Effectively,
we may use the number of levels contained in the Fermi block whose
entropy is maximal, $\ell_{max}$. We will now show that this magnitude
bears strong correlation with other possible estimates, based on two
very relevant magnitudes in a pairing system such as the {\em
  condensation energy} $E^C(L,g)$ (defined as the difference between
the GS energy and the energy of the corresponding Fermi sea), and the
{\em spectroscopic gap} $E^G(L,g)$ (defined as the excess energy
obtained when one pair is broken, blocking levels $\ell_F$ and
$\ell_F+1$).

\begin{figure}
\epsfig{file=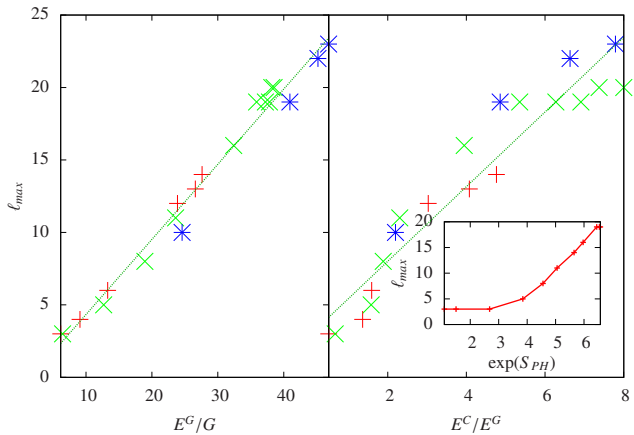,height=6cm}
\caption{\label{fig.swave.lmax}Left: Correlation between $E^G/G$ and
  $\ell_{max}$ in model (\ref{ham.richardson}), slope is $1/2$. Right:
  Correlation between $E^C/E^G$ and $\ell_{max}$, slope is $\approx
  2.5$. In both cases, $g\in [0.2,4]$ and $L=30$, $40$ and $50$
  (color). Inset: relation between the particle-hole entropy $S_{PH}$
  and $\ell_{max}$.}
\end{figure}

Figure (\ref{fig.swave.lmax}, left) shows that $\ell_{max}$ correlates
very strongly $E^G/G$, i.e.: the spectroscopic gap in units of the
dimensionful coupling constant $G=gd$. This correlation extends to a
wide range of values of $g\in [0.2,4]$, and for $L=30$, $40$ and
$50$. For very large coupling, the spectroscopic gap is roughly
proportional to $g$, i.e., the relation $E^G/G\approx L$ holds. Thus,
we postulate that, for all $g$, only the closest $L_{act}$ energy
levels to the Fermi energy are active, while all others are
frozen. Then, it is sensible to estimate $E^G/G\approx
L_{act}$. Indeed, the fit shows that $E^G/G\approx 2\ell_{max}$ with
very good accuracy. The reason for this factor 2 is {\em entanglement
  monogamy} \cite{Coffman.00}: if $\ell_{max}$ levels are strogly
entangled to the environment, it is reasonable to presume that they
are correlated to {\em other set} of $\ell_{max}$ levels. The same
correlation is observed when measuring the {\em no-broken-pairs} gap,
i.e.: the energy difference to the first excited state of the
hard-core boson Hamiltonian (\ref{ham.richardson}).

The right panel in figure (\ref{fig.swave.lmax}) shows the correlation
between $\ell_{max}$ and another estimate of the number of active
pairs: the condensation energy divided by the spectroscopic gap,
$E^C(L,g)/E^G(L,g)$. In BCS theory, the condensation energy is
proportional to the number of pairs times the gap. Thus, again, we
expect $E^C/E^G \approx L_{act}$. In this case, we obtain $E^C/E^G
\approx \ell_{max}/3$, with a worse fit. The inset in figure
(\ref{fig.swave.lmax}) shows the relation between $S_{PH}=S({\cal
  L}(\ell_F))$, the particle-hole entropy, and $\ell_{max}$. Indeed,
for large enough $\ell_{max}$ ($g>0.5$) there appears to be a {\em
  linear relation} between $\exp(S_{PH})$ and $\ell_{max}$. Indeed,
$\exp(S)$ measures the {\em number of relevant states} required to
describe a certain block, and this value is correlated with the number
of active pairs, as expressed in equation (\ref{richardson.gs.ginf}).

 
We now turn to the study of the entanglement properties of the 
1D $p$-wave analogue of the Richardson model studied in the
previous section whose Hamiltonian  is given by 
\cite{Ibanez.PRB.09,Dunning.JSTAT.10}:

\begin{equation}
H_p/d= \sum_j j^2\ b^\dagger_j b_j - g \sum_{j\neq j'}
j\cdot j'\ b^\dagger_jb_{j'} 
\label{ham.pwave}
\end{equation}
The phase diagram of this model can be divided into three regions:
weak coupling ($\frac{1}{g} \geq 1 - x$), weak pairing ($1 - 2 x <
\frac{1}{g} \leq 1 - x$) and strong pairing ($\frac{1}{g} < 1 - 2 x$),
where $g$ is the BCS coupling constant and $x= N_p/L$ is the filling
fraction. The weak pairing and strong coupling regions are separated
by the so called Read and Green line (RG) where the quasiparticle
energies are gapless \cite{Read_Green.PRB.00}. The corresponding phase
transition in 2D is of third order, as found by Rombouts and
collaborators \cite{Rombouts.PRB.10}, and is characterized by the
winding number of the BCS order parameter
\cite{Volovik.JETP.88,Read_Green.PRB.00}. The weak pairing phase,
being topologically non trivial, has a Pfaffian wave function similar
to the so called Moore-Read (MR) state of the FQH at filling fraction
$\nu = 5/2$ mentioned in the introduction. In reference
\cite{Ibanez.PRB.09,Dunning.JSTAT.10} it was found that the MR state
is the exact GS of the Hamiltonian (\ref{ham.pwave}) on the MR line
$\frac{1}{g} = 1 - x$. All the roots of the Richardson-like equations
vanish, and therefore the total energy of the state, given by the sum
of the roots, vanishes as well. The MR line corresponds to a zero
order phase transition meaning that the GS energy is discontinuous
when crossing it \cite{Dunning.JSTAT.10}. However all the remaining
observables does not exhibit any kind of jump. For this reason it is
generally believed that the weak pairing and weak coupling regions,
that are separated by the MR line, belong in fact to the same phase.

It is thus of great interest to study the entanglement spectrum and
entropy in connection with the nature of the quantum phase transitions
across the RG and MR lines. To this end we fix the filling fraction to
$x = 1/4$, so that the RG point is met at $g = 2$, while the MR point
is met at $g = 4/3$. 

The maximal entropy $S^F_{max}(L,g)$ of the Fermi blocks exhibits a
similar behavior to the $s$-wave case shown in (\ref{fig.swave.sl},
inset) and equation (\ref{slogl}): for large enough $g$ it grows as
$1/2\log(L) + \beta(g)$, with $\beta(g)$ attaining its maximum at the
RG point ($g=2$), signalling the divergence of the size of the Cooper
pairs \cite{Rombouts.PRB.10}. The maximal entropy of the low-energy
blocks, $S^L_{max}(L,g)$ does not show any special feature near that
point. Nonetheless, they take the same value at the MR point, $g=4/3$:
$S^F_{max}(L,g_{MR})=S^L_{max}(L,g_{MR})$.

\begin{figure}[hptb]
\epsfig{file=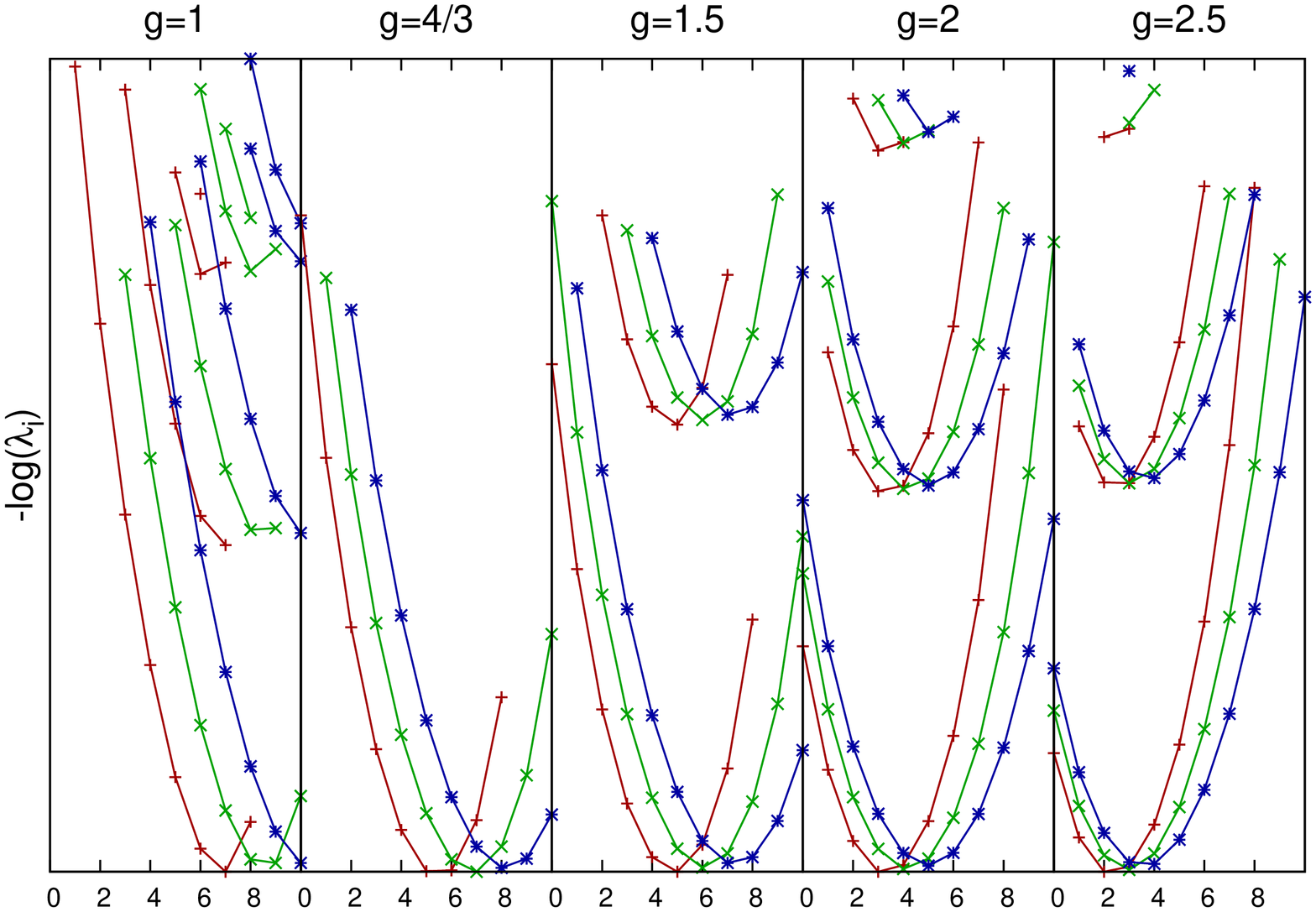,width=6cm,angle=270}
\epsfig{file=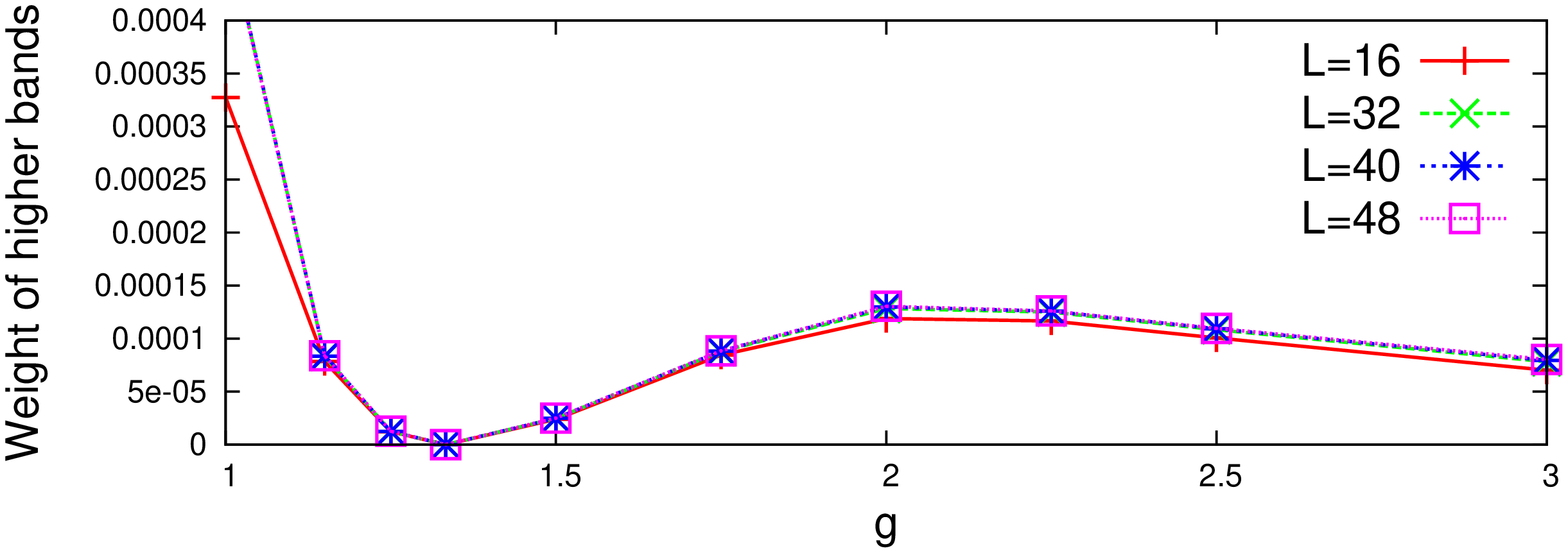,width=3cm,angle=270}
\caption{\label{fig.pwave.entsp}Top: Entanglement spectra of ${\cal
    L}(\ell_F)$ for the GS of (\ref{ham.pwave}). Bottom: weight of the
  higher bands as a function of $g$.}
\end{figure}

We have also obtained the entanglement spectrum of the particle-hole
block ${\cal L}(\ell_F)$ for $x=1/4$ and several values of $g$
belonging to different domains of the phase diagram. Figure
(\ref{fig.pwave.entsp}, top) shows the entanglement energies versus
the number of pairs below the Fermi point. As in the $s$-wave case, we
can recognize the parabolas which are responsible for the
$S_{max}\approx 1/2 \log(L)$ behavior. The vertices of the parabolas
shift leftwards as $g$ increases, from $\ell_{max}=\ell_F(=10)$, at
$g=0$ (all pairs are below the Fermi level) to
$\ell_{max}=x\ell_F(=2.5)$ as $g\to\infty$, because pairs become
equally distributed. A remarkable feature is that at $g=4/3$, i.e.:
the MR point, {\em the higher bands disappear}. The entanglement
spectrum consists only of the lower band, as far as the DMRG can
distinguish. This result agrees with the exact GS on the MR line which
is a perfect condensate of Cooper pairs
\cite{Ibanez.PRB.09,Dunning.JSTAT.10}. This explains the previous
observation that $S^F_{max}(L,g_{MR})=S^L_{max}(L,g_{MR})$. Figure
(\ref{fig.pwave.entsp}, bottom) depicts the weight of the higher bands
as a function of $g$ for a few values of $L$. The weight falls to zero
at $g=4/3$ (MR point), as explained above. Furthermore, it presents a
{\em local maximum} at $g=2$, the RG point, in accordance to the
divergence of the size of Cooper pairs \cite{Rombouts.PRB.10}.

To sum up, in this letter we have explored the entanglement of blocks
of energy-levels in the ground state of pairing Hamiltonians,
emphasizing the logarithmic behavior of the maximal entanglement
entropy, $1/2\log(L)$, which is related to the permutation-symmetry of
the state in the strong coupling limit, and the relation between the
size of the maximally entangled block and the number of active
pairs. Alongside, we have characterized the entanglement spectrum and
its parabolic band structure, clarifying the nature of the Moore-Read
and Read-Green points in the $p$-wave superconductivity phase diagram.

We thank J. Dukelsky, O. Castro-Alvaredo and B. Doyon for very useful
discussions, and acknowledge financial support from the Spanish
Government research projects FIS2012-33642 and the Severo Ochoa
program. J.R.-L. acknowledges also research project FIS2009-07277 and
ERC grant QUAGATUA.

\end{document}